\PassOptionsToPackage{table,xcdraw}{xcolor}
\documentclass[sigconf,screen]{acmart}
\usepackage{tabularx}
\usepackage{algorithm}
\usepackage{algpseudocode}
\usepackage{amsmath,amsfonts}
\usepackage{graphicx}
\usepackage{textcomp}
\usepackage{xcolor}

\usepackage{threeparttable}
\usepackage{enumitem}
\usepackage{array}
\usepackage{wrapfig}
\usepackage{amsmath,amsfonts}
\usepackage{graphicx}
\usepackage{textcomp}
\usepackage{float}
\usepackage{listings}
\usepackage{xspace}
\usepackage{multirow}
\usepackage{amsthm}

\usepackage{balance}
\usepackage{algorithm}
\usepackage{colortbl}
\usepackage{subcaption}

\usepackage[skins]{tcolorbox}
\usepackage{xcolor}
\usepackage{multicol}
\usepackage{soul}
\usepackage{framed}
\usepackage{booktabs}
\usepackage{booktabs}
\usepackage{multirow}
\usepackage{siunitx}

\usepackage{adjustbox}    
\usepackage{xcolor}       

\newcommand*\colourcheck[1]{%
	\expandafter\newcommand\csname #1check\endcsname{\textcolor{#1}{\ding{52}}}%
}

\colourcheck{blue}
\colourcheck{green}
\colourcheck{red}
\newtcolorbox{boxB}[2][]{%
  enhanced,colback=white,colframe=black,coltitle=black,
  sharp corners,
  toprule=1.0pt,
  rightrule=0.3pt,
  leftrule=0pt,
  bottomrule=0pt,
  fonttitle=\itshape\scshape\large,
  left=0pt,right=5pt,top=5pt,bottom=3pt,
  attach boxed title to top right={yshift=-0.3\baselineskip-0.4pt,xshift=-5mm},
  boxed title style={tile,size=minimal,left=0.2mm,right=0.5mm,
    colback=white,before upper=\strut},
  title=#2,#1
}

\definecolor{sh_string}{rgb}{0.1, 0.6, 0.1}

\lstdefinelanguage{json}{
  basicstyle=\ttfamily,
  showstringspaces=false,
  morestring=[b]",
  stringstyle=\color{sh_string}
}

\newcommand{\tool}{\textsc{TestWeaver}\xspace}

\definecolor{royalblue}{rgb}{0.2549, 0.4118, 0.8824}
\definecolor{lightgray}{rgb}{211, 211, 211}
\definecolor{gray}{rgb}{0.5,0.5,0.5}

\newtcolorbox{promptbox}[1][]{ 
  colback=lightgray!10,  
  colframe=gray,      
  fonttitle=\bfseries,    
  fontupper=\scriptsize,
  title={\centering\scriptsize #1},
  before=\definecolor{royalblue}{rgb}{0.2549, 0.4118, 0.8824} 
}

\newtcolorbox{outputbox}[1]{ 
  colback=lightgray!10,  
  colframe=gray,      
  fonttitle=\bfseries,    
  fontupper=\scriptsize,
 before=\definecolor{royalblue}{rgb}{0.2549, 0.4118, 0.8824} 
}

\definecolor{custom-blue}{rgb}{0,0,0}

\newboolean{showcomments}
\setboolean{showcomments}{true}
\ifthenelse{\boolean{showcomments}}
 { \newcommand{\mynote}[2]{
      \fbox{\bfseries\sffamily\scriptsize#1}
        {\small$\blacktriangleright$\textsf{\emph{#2}}$\blacktriangleleft$}}}
        { \newcommand{\mynote}[2]{}}

\newcolumntype{L}[1]{>{\raggedright\arraybackslash}p{#1}}
\newtheorem{Observation}{Observation}
\newtheorem{Key Idea}{Key Idea}

\newcommand{\code}[1]{{\footnotesize\texttt{#1}}}

\definecolor{dkgreen}{rgb}{0,0.6,0}
\definecolor{gray}{rgb}{0.5,0.5,0.5}
\definecolor{lightgray}{rgb}{211, 211, 211}
\definecolor{mauve}{rgb}{0.58,0,0.82}

\definecolor{c1}{HTML}{f4cccc}
\definecolor{c2}{HTML}{f5cdcd}
\definecolor{c3}{HTML}{fffcfc}
\definecolor{c4}{HTML}{ffffff}
\definecolor{c5}{HTML}{ffffff}
\definecolor{c6}{HTML}{fffdfd}
\definecolor{c7}{HTML}{f5cfcf}
\definecolor{c8}{HTML}{fffbfb}
\definecolor{c9}{HTML}{ffffff}
\definecolor{c10}{HTML}{fffdfd}
\definecolor{c11}{HTML}{fefafa}
\definecolor{c12}{HTML}{fef7f7}
\definecolor{c13}{HTML}{ffffff}
\definecolor{c14}{HTML}{fffefe}
\definecolor{c15}{HTML}{ffffff}
\definecolor{c16}{HTML}{fefafa}
\definecolor{c17}{HTML}{fdf3f3}
\definecolor{c18}{HTML}{fffefe}
\definecolor{c19}{HTML}{fdf5f5}
\definecolor{c20}{HTML}{ffffff}

\copyrightyear{2026}
\acmYear{2026}
\setcopyright{cc}
\setcctype{by-nc-nd}
\acmConference[ICSE '26]{2026 IEEE/ACM 48th International Conference on Software Engineering}{April 12--18, 2026}{Rio de Janeiro, Brazil}
\acmBooktitle{2026 IEEE/ACM 48th International Conference on Software Engineering (ICSE '26), April 12--18, 2026, Rio de Janeiro, Brazil}
\acmPrice{}
\acmDOI{10.1145/3744916.3787805}
\acmISBN{979-8-4007-2025-3/2026/04}

\begin{document}

\title{{\tool}: Execution-aware, Feedback-driven Regression Testing Generation with Large Language Models}

\author{Cuong Chi Le}
\orcid{0009-0006-1322-7396}
\affiliation{
	\institution{FPT Software AI Center}
	\city{Hanoi}
	\country{Vietnam}}
\email{cuonglc4@fpt.com}

\author{Cuong Duc Van}
\orcid{0009-0008-6700-1477}
\affiliation{
	\institution{FPT Software AI Center}
	\city{Hanoi}
	\country{Vietnam}}
\email{cuongvd17@fpt.com}

\author{Tung Duy Vu}
\orcid{0009-0003-0532-1428}
\affiliation{
	\institution{VinUniversity}
	\city{Hanoi}
	\country{Vietnam}}
\email{21tung.vd@vinuni.edu.vn}

\author{Minh V.T. Pham}
\orcid{0000-0002-6457-1123}
\affiliation{
	\institution{FPT Software AI Center}
	\city{Hanoi}
	\country{Vietnam}}
\email{minhpvt@fpt.com}

\author{Hoang N. Phan}
\orcid{0009-0008-1197-8115}
\affiliation{
	\institution{Nanyang Tech. University}
	\country{Singapore}}
\email{c210055@e.ntu.edu.sg}

\author{Huy N. Phan}
\orcid{0000-0003-1462-9515}
\affiliation{
	\institution{FPT Software AI Center}
	\city{Hanoi}
	\country{Vietnam}}
\email{huypn16@fpt.com}

\author{Tien N. Nguyen}
\orcid{0009-0006-7962-6090}
\affiliation{
	\institution{Univ. of Texas at Dallas}
	\city{Dallas}
	\country{USA}}
\email{tien.n.nguyen@utdallas.edu}

\begin{abstract}
While recent advances in large language models (LLMs) have shown promise in automating test generation for regression testing, they often suffer from limited reasoning about program execution, resulting in stagnated coverage growth—a phenomenon known as the coverage plateau. 
This paper presents {\tool}, a novel LLM-based approach that integrates lightweight program analysis to create a focused execution context that assists LLMs in better test generation.
{\tool} strategically chooses the following components to overcome LLMs’ limited~reasoning on complex execution: (1) it reduces hallucinations and improves focus by supplying the LLM with the backward slice from the target line instead of a full program context; (2) it identifies and incorporates close test cases—those that share control-flow similarities with the path to the target line—to provide focused execution context within the LLM’s context window; and (3) it enhances LLM's reasoning with execution in-line annotations that encode variable states as comments along the executed path. By equipping LLMs with these targeted and contextualized inputs, it improves coverage-guided test generation and mitigates redundant explorations. Empirical results show that {\tool} accelerates code coverage growth and generates more effective test cases than the state-of-the-art~approaches.
\end{abstract}



\maketitle

\section{Introduction}
\label{sec:intro}

Regression testing is a technique used to ensure that future code changes do not break the existing functionality of the current software version. It involves creating and preserving test cases that can be re-executed to validate the system’s behavior as it evolves.

Creating a regression test case can begin with the current version of the code that is known to function correctly, using its existing behavior as a baseline for future validation. By capturing how the code behaves now, we can detect unintended changes introduced by future modifications. To strengthen the effectiveness of regression testing, it is important to {\em add test cases that cover a broader range of scenarios and execution paths}. This expanded coverage increases the likelihood of uncovering hidden bugs and ensures more comprehensive protection against regressions.

Recently, several researchers have explored the use of machine learning (ML) and large language models (LLMs) for automated test generation~\cite{pizzorno2025coverup} and fuzzing in the context of regression testing~\cite{fuzz4all,lemieux2023codemosa}. While having promising results, they largely depend on the reasoning capabilities of LLMs to generate test cases that exercise specific lines of code, conditions, or execution paths in the target program. Most existing techniques rely heavily on prompt-based querying of LLMs, which—according to a recent study by Chen {\em et al.}~\cite{chen2025reasoning}—struggle with accurately reasoning about program execution. This limitation arises because LLMs are primarily trained on static code representations, lacking explicit runtime context or behavioral data. As a result, their ability to simulate execution, predict program states or code coverage is inherently limited.

In the context of coverage-guided test generation, one common consequence of this limitation is the phenomenon known as the {\em coverage plateau}: as test generation iterations continue, the rate of discovering new execution paths diminishes. For the LLM-based approaches, LLMs tend to produce test cases that fail to increase coverage, often generating different inputs yet covering the already-covered lines of code. Current approaches typically address this by simply {\color{custom-blue}{{\em passive (re-)prompting}}} the LLMs {\em without offering helpful guidance}, which exacerbates the stagnation. This phenomenon has been observed and studied in prior works on coverage-guided test generation~\cite{lemieux2023codemosa,pizzorno2025coverup}, where coverage grows slowly over time.

In this paper, we propose {\tool}, a coverage-guided, LLM-based approach for automated regression test generation that integrates lightweight program analysis (PA) to create a {\color{custom-blue}{focused~execution context that assists LLMs}} in better test generation.
The goal is to accelerate coverage growth by guiding LLMs toward generating higher-quality test cases that exercise previously uncovered lines of code. {\tool} is built upon the integration of our core ideas:

First, instead of feeding the entire program—which may be long and complex—into the LLM, we extract and provide only the {\bf backward slice} from the target line of interest. This slice includes the program statements that have a potential influence on the execution of that line. By narrowing the input to only the relevant parts of the code, we reduce the chance of hallucinations and minimize error propagation, while enabling the LLM to focus on the critical execution logic that affects the target line’s reachability.


Second, instead of re-prompting the LLM to generate a new test case for all the not-yet-covered target lines (which may even require conflicting execution paths with one single test case), {\tool} provides an {\bf execution-aware, constructive feedback}. It leverages the {\em execution context from the current test suite}. The insight is that {\em the uncovered line might already be near the execution path of an existing test case}, even if not covered directly. Thus, giving the LLM the knowledge of execution behavior across the test suite offers a more informed basis to reason on what remains to be covered.

However, providing full execution information for all test cases is impractical due to LLM context window limitations. To address this, we draw inspiration from concolic testing~\cite{sen2005cute,godefroid2005dart} and propose selecting {\bf a "close" test case} to guide the LLM. Specifically, a test case $T$ is close to a target line $l_t$ if: 
1) it executes a control-dependent condition of $l_t$; and 2) among such conditions, it covers the one that is closest in line distance to $l_t$. This approach exploits control dependence chains and execution traces to guide the retrieval structurally.
This targeted strategy ensures that the LLM is grounded in relevant dynamic context, increasing the likelihood that the new test case will cover new code and avoid redundant explorations.

Finally, in each test generation iteration, we further equip the LLM with {\bf execution in-line annotations}—a representation of dynamic program states. These annotations embed the values of relevant variables as in-line comments for each executed line at the key points during program's execution. This {\em execution-aware contextualization} allows the LLM to observe subtle data and control dependencies, {\em enabling it to better infer the values needed to steer execution along specific paths and ultimately reach the target line}.

We conducted an extensive empirical evaluation of {\tool}.
Using the CodaMosa benchmark—which includes 35 real-world Python projects and over 100{,}000 lines of code—we compared \tool against two state-of-the-art baselines: \textsc{CodaMosa}~\cite{lemieux2023codemosa} and \textsc{CoverUp}~\cite{pizzorno2025coverup}. \tool achieves \textbf{68\%} line coverage and \textbf{62\%} branch coverage, outperforming \textsc{CoverUp} (61\% / 47\%) and \textsc{CodaMosa} (46\% / 23\%). It generalizes well to highly-complex and longer programs, maintaining strong coverage and avoiding the coverage plateau common in the baselines by providing more constructive and execution-aware context to the LLM after each retry. 
Our tool also strikes a good balance between effectiveness and cost. It reaches the highest coverage point in 2.76$\times$ faster than \textsc{CodaMosa} while achieving higher coverage (68\% vs 46\%). In about the same time, it reaches higher coverage than \textsc{CoverUp} (68\% vs 61\%), with less token and money costs.
Ablation study shows that removing any core component (backward slicing, closest-test retrieval, or execution in-lines) reduces coverage by up to 12\%, confirming their critical role. 
In brief, we make the following contributions:
\begin{enumerate}[leftmargin=*]
    \item \textbf{\tool: A Coverage-Guided Test Generation Framework} {\color{custom-blue}{that integrates three ideas: (1) slicing shortens context by removing irrelevant code; (2) closest test cases enable contrastive reasoning, increasing the likelihood of covering new code while avoiding redundant exploration; and (3) execution paths and values provide richer feedback than simple coverage.}}
    

    \item {\bf Focused Execution-Aware Context}: {\color{custom-blue}{This feedback mechanism retrieves the most relevant test case and uses execution inlining to help the LLM better reason on program behaviors, and effectively generate test cases covering the target code.}}

    

    \item \textbf{Extensive Empirical Evaluation} We evaluate our tool on 35 Python projects against state-of-the-art baselines, showing superior code coverage, robust performance on complex code with less coverage plateau, and efficiency in runtime and token~cost. 

    
\end{enumerate}

\section{Motivation}
\label{sec:motivation}

\subsection{Example and Observations}

\begin{figure}[t]
\begin{lstlisting}[language=Python, basicstyle=\footnotesize\ttfamily]
def evaluate_sequence(arr: list[int]):
    score = 0
    freq = {}
    min_val, max_val = min(arr), max(arr)

    for x in arr:
        freq[x] = freq.get(x, 0) + 1
        score += x
    if len(arr) < 4:
        if arr[0] % 2 == 0:
            score += 4
        else:
            score -= 2
            if arr[-1] == 3:
                score += 10
    elif score % 5 != 0:
        if max_val - min_val > 50:
            score += 5
            if sum(1 for x in arr if x % 2 == 0) > len(arr) // 2:
                score *= 2
        elif 0 in freq:
            score += 7
    else:
        if arr == sorted(arr):
            score += 3
        elif all(v < 3 for v in freq.values()):
            score -= 1 # target line
        else:
            score += 6

    return score
\end{lstlisting}
\vspace{-15pt}
\caption{LLMs struggle with execution reasoning: an example of GPT-4o's generating a test case to cover line 27.}
\label{fig:motiv1}
\end{figure}

\begin{figure*}[h!]
  \centering
  \begin{adjustbox}{center}
    \begin{minipage}[c]{0.4\textwidth}
      \centering
      \includegraphics[width=\textwidth]{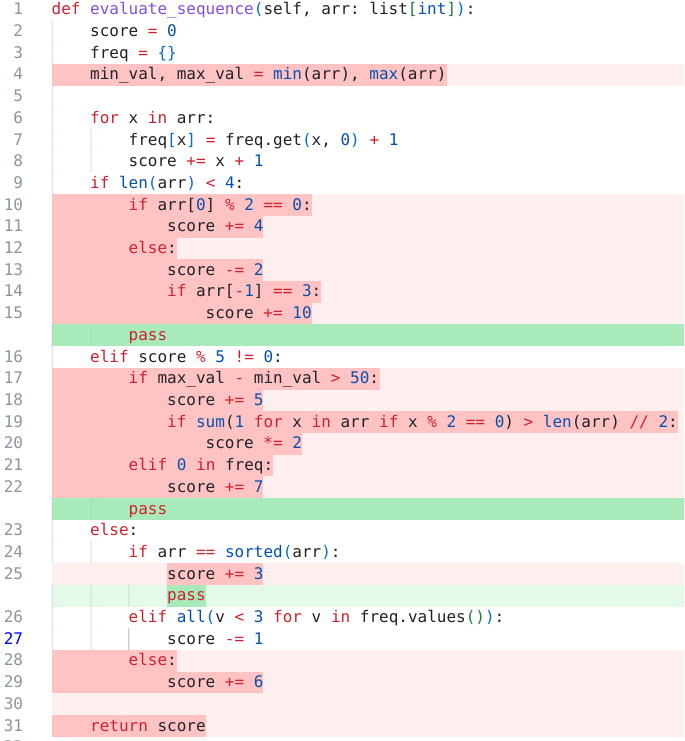}
      \vspace{-1mm}
      
      {\sffamily\small \textcolor{red}{Correct Test Case: 9/100\\Acc@1: 9\%}}
    \end{minipage}
    \hspace{0.25em}
    \begin{minipage}[c]{0.06\textwidth}
      \centering
      \includegraphics[width=\textwidth]{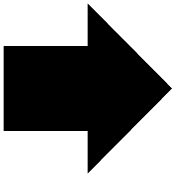}
    \end{minipage}
    \hspace{0.25em}
    \begin{minipage}[c]{0.4\textwidth}
      \centering
      \includegraphics[width=\textwidth]{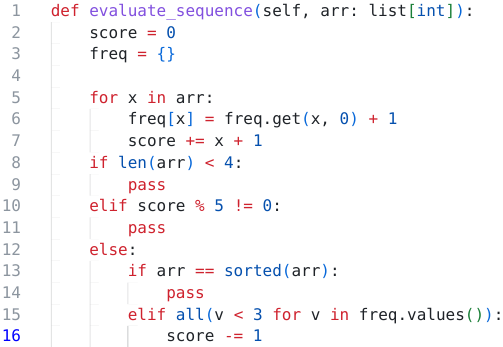}
      \vspace{1mm}
      
      {\sffamily\small \textcolor{teal}{Correct Test Case: 100/100\\Acc@1: 100\%}}
    \end{minipage}
  \end{adjustbox}
  \vspace{-12pt}
  \caption{Improvement in test case generation using dynamic backward slicing}
  \label{fig:motiv2}
\end{figure*}

\subsubsection{LLMs Struggle with Execution Reasoning}
Let us use an example {\color{custom-blue}{in Figure~\ref{fig:motiv1}}} to illustrate the issues and motivate our approach. In automated test generation for regression testing, the goal is to augment the existing test suite with new test cases that exercise a broader range of code lines and execution paths. Approaches leveraging large language models (LLMs) for this task must effectively guide the model to reason about program execution behavior. However, program execution prediction—also referred to as dynamic code reasoning—remains a significant challenge for LLMs~\cite{chen2025reasoning,codeflow}.

To illustrate this point, we examine a Python function that computes a score based on the input array \code{arr}. We prompted GPT-4o to generate a test case specifically targeting line 27 and explain the code coverage, i.e., which lines have been covered for that generated test case. Execution of this line depends on several dynamic conditions, including the length of the array, whether the total sum is even, and whether the score exceeds a threshold of 10. Satisfying these conditions requires navigating through nested loops and conditionals—an inherently complex reasoning task for automated test generation. Despite the seemingly straightforward prompt, GPT-4o failed to produce any test case that would trigger line 27. 


\begin{Observation} [LLMs Struggle with Execution Reasoning]
{\color{custom-blue}{While LLMs generally perform well in several code-related tasks}}, they often struggle to reason about \emph{actual code execution} and the dynamic evolution of the program state.
%
\end{Observation}


In the example, the value of \code{score} evolves across multiple loop iterations and branches. Correctly generating a test case that reaches line 27 (\code{score -= 1}) requires reasoning about how this variable changes throughout execution. Since LLMs primarily rely on static syntax rather than execution semantics, they often overlook the interplay between conditions and runtime state.

To illustrate this, we prompted GPT-4o to generate a test case that would cover line 27. It produced the input \code{arr = [2, 3, 4, 6, 7, 9]} and explained that execution would skip line 16 (due to a miscalculation that \code{score = 19} satisfies \code{score \% 5 == 0}) and line 24 (claiming the array is not sorted), which would supposedly lead to line 27 executing. However, in reality, the program did enter line 16, triggering line 19, which doubled \code{score} and changed the execution path, skipping line 27 entirely. GPT-4o’s explanation reveals an incorrect mental model of control flow and variable states, highlighting its difficulty in reliably predicting dynamic execution, even in simple cases.


Li {\em et al.}~\cite{yi2025blended} have reported that one contributing factor to the suboptimal performance of LLMs in code execution reasoning is their tendency to hallucinate when processing long or complex source code. To address this, we applied backward program slicing starting from line 27 and extracted the relevant statements that influence the program state at line 27. We then provided this slice to GPT-4o (Figure~\ref{fig:motiv2}). As a result, GPT-4o successfully generated a test case which exercises that line. Moreover, as shown in Figure~\ref{fig:motiv2}, when running multiple times, GPT-4o was able to produce 100 out of 100 test cases that cover line 27 using the sliced code, compared to only 9 out of 100 on the original complete program.


\begin{Observation}[Slicing Helps LLMs Generate Better Test Cases]
Program slicing helps focus the LLM's attention on the critical statements that influence a specific target line. This reduces~distractions from unrelated code—such as loop constructs—and improves the LLM’s precision in generating test cases covering a target line.
\end{Observation}

The backward slicing isolates the \code{if} conditions around line~27, guiding the LLM to generate test cases more likely to cover line~27.

\begin{figure*}[t]
  \centering
  \includegraphics[width=\linewidth]{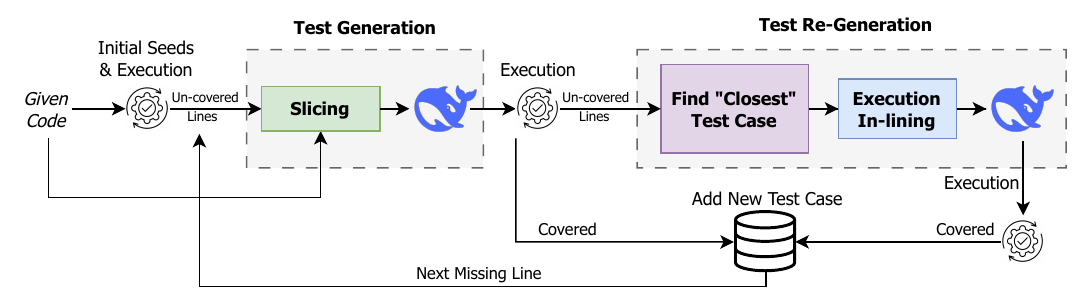}
  \vspace{-21pt}
  \caption{Overview of the {\tool} pipeline. The process begins by generating initial test seeds and identifying uncovered lines through execution. For each uncovered line, a backward slice is computed to produce a sliced code, which is then provided to the LLM for test generation. If the generated test fails to cover the target, the re-generation stage is triggered: the closest failing test is retrieved from the current suite, its execution trace is used to annotate the sliced code with variable values (execution in-lines), and the enriched prompt is re-submitted to the LLM. Any successful test is added to the suite, and the process continues until all uncovered lines have been addressed or a time limit is reached.
}
  \label{fig:overview}
\end{figure*}




\subsubsection{Coverage Plateau}
During the iterative process of generating new test cases to improve code coverage, LLMs typically identify many new execution paths early on, leading to a rapid increase in code coverage. However, as the process continues, the rate of discovering new paths declines. The remaining unexplored paths tend to be more complex, often requiring specific and rare input conditions to be triggered. As a result, coverage growth slows significantly, leading to a phenomenon known as the coverage plateau~\cite{lemieux2023codemosa,pizzorno2025coverup}.

The limited reasoning ability of LLMs for dynamic program behavior is one cause of the coverage plateau problem. Another key factor is that existing LLM-based test generation approaches~\cite{pizzorno2025coverup} do not provide constructive feedback to help the LLMs. {\color{custom-blue}{Instead, they repeatedly prompt the LLM to cover all remaining uncovered lines $L$ {\em without offering any guidance to help the model improve in the next attempt}—a strategy we call {\bf passive feedback}. This passive approach has critical shortcomings. Providing the entire set of uncovered lines $L$ can easily confuse the LLM, since covering multiple lines may require conflicting execution paths within a single test case. Simply repeating the same coverage request without any context often fails to help the LLM make progress, directly contributing to stagnation in coverage growth. This passive feedback does not help the LLMs recognize the nuances in the execution with different test cases, in which LLMs by themselves have limited capability.}}



\begin{Observation}[Coverage Plateau]
The code coverage plateau problem is caused by the limited capability of LLMs in dynamic code reasoning and the passive feedback strategy in the state-of-the-art, LLM-based regression test case generation approaches.
\end{Observation}


We hypothesize that {\em providing execution data—such as execution traces or symbolic execution results}—can significantly improve an LLM’s understanding of dynamic program behavior, enabling more effective and targeted test case generation. For example, dynamic symbolic execution can produce path constraints that describe how to reach a target statement, as in concolic testing~\cite{sen2005cute,godefroid2005dart}.~Yet,~solving these constraints, especially when they involve multiple data types, is inefficient or even infeasible for SMT solvers. On the other hand,
supplying execution information for the entire test suite and its covered lines will exceed the LLM’s context window limits. 

\begin{Observation}[Execution-aware Feedback]
Providing an LLM with execution information on the current test suite could help it generate test cases for uncovered lines. 
\end{Observation}


\subsection{Key Ideas}


From the observations, {\tool} is designed with the following:

\subsubsection{Key Idea 1. [{\bf Program Slicing} for Targeted Test Case Generation]}
\label{sec:key-idea-1}
Program slicing reduces code complexity by isolating the relevant portions that influence a specific target line. By extracting only the essential information needed for test case generation, slicing helps the LLM concentrate on the most critical parts of the code. This focused context minimizes the risk of hallucination and ensures that generated test cases align more closely with the actual execution flow. Additionally, slicing narrows the search space, making the LLMs to generate test cases more targeted and efficient.

\subsubsection{Key Idea 2. [{\bf "Closest" Test Selection} to the Target Line]}
We aim to leverage the coverage information from the entire test suite to guide test case generation. However, encoding all test cases along with their covered lines and associated program states exceed the context window limitations of LLMs. To address this, we propose a strategy that uses the full test suite to identify a subset of similar test cases—those that cover execution paths "closest" to the target line (Section~\ref{subsubsec:retrieve}). By focusing on this similarity-based subset, we can guide the LLM to generate new test cases that meaningfully extend coverage without duplicating prior explorations. This approach mitigates the risk of out-of-context generation or hallucination and allows for more precise and efficient targeting of uncovered lines.

\subsubsection{Key Idea 3. [Enhancing Code Execution Understanding with {\bf Inlining Execution Data}]}  
To enhance an LLM's ability to reason about code execution and generate test cases that cover specific lines of code, we integrate the LLM with external tools to support in-line execution feedback. Specifically, after executing a generated test case, we annotate the corresponding lines with the observed program states including the runtime values of the variables involving in the statements as in NeXT~\cite{icml2024next}. This enables the LLM to track the variable states and data flows in "execution". Inspired by agent-based systems, this approach allows the LLM to simulate actual program execution more effectively, resulting in more accurate and reliable generation of test cases to cover the target line.




\vspace{-3pt}
\section{{\tool} Test Generation Workflow}
\label{sec:method}


\subsection{Overview}

Our pipeline (Figure~\ref{fig:overview}) for automated test generation begins by producing an initial seed test suite, {\color{custom-blue}{which can be generated by any initial corpus test generation algorithm~\cite{herrera2021seed,pailoor2018moonshine}. In our experiment, we leverage LLM to generate $N$ initial test cases.}} Then, all the lines that remain uncovered are identified. These uncovered lines then drive the slice‐guided test generation step: for each target line, a backward slice (Section~\ref{sec:program-slicing}) is computed to isolate the minimal fragment of source code that influences its execution. That fragment—together with the target line itself—is presented to a LLM, whose attention is thus focused solely on the relevant control and data dependencies when generating a new test case.
{\color{custom-blue}{
Specifically, when the generated candidates still fail to cover the target line after a predefined number of attempts, our tool invokes the execution‐inlining test regeneration step. A heuristic retrieval method selects from the existing suite the ``closest'' test case (Section~\ref{subsubsec:retrieve}) whose execution trace passes through the nearest conditional statement of the target line but thereafter diverges along a different path. The backward slice of that ``closest'' test case is then instrumented with the inline annotations of runtime values observed under that test case,  providing the LLM with concrete execution contexts. These enriched slices are iteratively presented to the LLM  until the target line is covered or a predefined number of iterations is reached. Any successful test is added to the test suite, and its coverage is subsequently used to address the remaining uncovered lines.
}}



\vspace{-2pt}
\subsection{Test Generation with Program Slicing}
\label{sec:program-slicing}

{\color{custom-blue}{

After identifying a target line of code to be covered, {\tool} computes a dynamic backward slice to keep only the minimal number of statements that influence the execution
of the target line. Because there does not exist any dynamic program slicing library for Python code that fits our workflow, we {\em re-implemented} Agrawal and Horgan’s dynamic backward slicing algorithm (Approach 2)~\cite{agrawal1990dynamic}. Additionally, because that algorithm supports only intra-procedural slicing, we also {\em extended it to perform inter-procedural backward slicing} through multiple involved functions. Let us detail it.

In the first phase, {\tool} 
builds the static program dependence graphs (PDGs) for all functions and module-level statements encountered. Next, to extend Agrawal and Horgan's algorithm to support inter-procedural flows, it constructs a system dependence graph (SDG) by linking these PDGs of those functions via the call sites (and actual arguments), the function entries (and formal arguments), \code{return} statements, and variable assignments and their~sources. Initially, all the SDG nodes (statements) and edges (program dependencies) are unmarked (i.e., not-executed). In the next step of this phase, following the execution of the given test case, our algorithm marks those SDG nodes and edges whose dependencies arise as executed including the inter-procedural dependencies.

In the second phase, i.e., the slicing phase, from the target statement, the algorithm traverses the SDG backward only along the marked nodes and edges (i.e., the executed statements and actual control and data dependencies, including match-case patterns). This helps {\tool} include only {\em the statements that were executed and came from the executed path}. Specifically, in addition to the data and control dependencies, the algorithm follows different types of relations among variables and references~\cite{yi2025blended}: (1) \code{def-use} (definition to reference): a definition and a reference to a variable have a \code{def-use} relation if there exists a control flow from the statement containing that definition to the statement containing that reference without~any intervening redefinition of that variable; and~(2)~\code{info-flow} (reference to definition): for a given statement $S$, a reference to a variable $v_1$ has an info-flow relation with the definition of another variable $v_2$ if the value of the variable $v_1$ on the entry to the execution of the statement $S$ may affect the value of $v_2$ on the exit from the statement $S$. The nodes (i.e., statements) that have been reached via these relations are collected into the resulting dynamic backward slice from the target statement.

Note that, during this traversal of the slicing phase, the algorithm also propagates/expands the slice through the actual function calls and returns in order to perform inter-procedural slicing.

This program slice and the target line are presented to the LLM with the test generation prompt in Table~\ref{tab:prompt_templates}.1. This guides the LLM to focus on relevant statements and their dependencies to the target line to generate a test case covering it. If the generated test case from the LLM is un-executable due to compile errors, we use the same program analysis in \textsc{CoverUp} to guide the LLM in fixing it. After a limited number of attempts of fixing, we drop that test case.

}}

\subsection{Retrieving ``Closest'' Test Case to Target Line}
\label{subsubsec:retrieve}

\subsubsection{Formulation}

Given a target line $l_t$ in the program that remains uncovered, we aim to identify a test case $T$ from the current test suite $\mathcal{T}$ that has traversed the code \emph{as "closest" as possible} to $l_t$. Instead of relying on full execution path comparison, we leverage the \emph{control dependencies} of the line $l_t$, specifically, the conditional constructs (e.g., \code{if}, \code{while}) needed to be satisfied for $l_t$ to execute.~Let:

\begin{itemize}
  \item $\mathsf{Cond}(l_t)$ = the ordered list of control-dependent conditional statements governing $l_t$, sorted by proximity (in source line number) to $l_t$
  \item $\mathsf{Exec}(T)$ = the set of lines executed by test case $T$
\end{itemize}

Then, a test case $T$ is considered \emph{close to $l_t$} if it satisfies:

\begin{enumerate}
  \item \textbf{Conditional Coverage Proximity}: The test case $T$ is considered close to $l_t$ if it executes a condition statement $c \in \mathsf{Cond}(l_t)$ such that the line distance $|c - l_t|$ is minimized:
  $$
  \text{Closeness}(T, l_t) = \min_{c \in \mathsf{Cond}(l_t) \cap \mathsf{Exec}(T)} |c - l_t|
  $$

  \item \textbf{Structural Control Dependency}: Only the condition statements that appear in the control flow of $l_t$ are considered. This ensures that the test case follows an execution path that nearly reaches the target line $l_t$-diverging only at its closest condition statement—provides valuable execution context for the LLM to generate a test case covering $l_t$.

\end{enumerate}

In summary, a test case $T$ is \textbf{close} to a target line $l_t$ if it executes a control-dependent condition of $l_t$, and among such conditions, it covers the condition that is closest in line distance to $l_t$. This approach exploits control dependence chains and execution traces to guide the retrieval structurally.
  

\subsubsection{Algorithm}

\begin{algorithm}[t]
\caption{{\color{custom-blue}{FindClosestTestCase($\mathcal{T}$, $l_t$, CDG)}}}
\label{alg:closest-test}
\begin{algorithmic}[1]
\Require Test suite $\mathcal{T}$, target line $l_t$, control dependency graph CDG
\Ensure Closest test case $T^* \in \mathcal{T}$ to target line $l_t$

\State $\mathsf{Cond}(l_t) \gets$ \textsc{GetControlConditions}($l_t$, CDG)
\State $T^* \gets \text{None}$, \quad $\delta_{\min} \gets \infty$

\ForAll{test case $T$ in $\mathcal{T}$}
  \State $E_T \gets$ \textsc{GetExecutedLines}($T$)
  \ForAll{$c \in \mathsf{Cond}(l_t)$}
    \If{$c \in E_T$}
      \State $\delta \gets |l_c - l_t|$
      \If{$\delta < \delta_{\min}$}
        \State $T^* \gets T$
        \State $\delta_{\min} \gets \delta$
      \EndIf
      \State \textbf{break} \Comment{Only consider closest conditional covered}
    \EndIf
  \EndFor
\EndFor

\State \Return $T^*$

\medskip
\Statex \textbf{Helper Functions:}
\Statex \textit{GetControlConditions}($l_t$, CDG): Returns the ordered list of conditions controlling $l_t$, based on the CDG.
\Statex \textit{GetExecutedLines}($T$): Returns the set of all program lines executed by test case $T$.

\end{algorithmic}

\end{algorithm}

Algorithm~\ref{alg:closest-test} shows our retrieval strategy based on conditional coverage proximity. It begins by identifying the set of control-dependent conditions for the target line $l_t$ using static analysis over the control dependencies (line 1).

We initialize the closest test case $T^*$ as \code{None}, and set the best (minimal) line distance $\delta_{\min}$ to $\infty$ (line 2). We iterate over each~test case $T$ in the current suite $\mathcal{T}$ (line 3). For each test case, its dynamic execution trace $E_T$ is obtained (line 4), which is the set of lines~covered during the execution. We then loop through each condition~$c$ that controls $l_t$ (line 5). If a condition $c$ is executed by $T$ (line 6), {\color{custom-blue}{the algorithm computes the absolute line difference $\delta = |l_c - l_t|$ (line 7)}}. If this $\delta$ is smaller than the current best distance $\delta_{\min}$ (line 8), we update $T^*$ as the new closest test case and record $\delta$ as the new minimum (lines 9–10). 
Since the conditions are ordered by proximity to $l_t$, we break early after the first match is found (line~12).

{\color{custom-blue}{
The \code{GetControlCondition()} function (Algorithm~\ref{alg:closest-test}) and the dynamic backward slicing algorithm support the following control-flow constructs in Python: \code{if}, \code{while}, \code{for}, \code{try-catch}, \code{with}, \code{try-except}, \code{finally}, \code{match-case}, \code{break}, \code{continue}, and \code{return}.
}}

\subsection{Test Re-generation with Execution In-lines}

Only after extracting the program slice on the closest test case relevant to a given uncovered line, we augment it with execution information in the form of \textit{execution in-lines}. Note that the initial program slice from the Test Generation module (Figure~\ref{fig:overview}) does not contain execution in-lines.
They are the annotations that explicitly encode the runtime values of variables at each statement, thereby exposing the concrete program state to the LLM. The annotations also encode the execution index of a statement when it was visited (i.e., the $k$-th statement executed).
The goal is to help the LLM understand how the execution flows under a test case. We expect that to help the LLMs to derive how it might need to change the input to cover the target line.

To obtain these values, the closest test case is first identified from the current test suite using Algorithm~\ref{alg:closest-test}. We execute the closest test case and the dynamic backward slice  from the target line is obtained. Then, we use a tool to record the execution information and instrument the backward slice with \textit{execution in-lines}. That is, the inline comments are inserted after each executable line to show the values of in-scope variables at the corresponding statement. The format of each in-line annotation is as follows:
\begin{lstlisting}[language=Python]
# (k) var1 = v1; var2 = v2; ...
\end{lstlisting}
Here, $(k)$ denotes the execution index, and each \code{var}=\code{value} pair~reflects the variable's value after executing that line. For example,
\begin{lstlisting}[language=Python]
x = a + b     # (1) a = 2; b = 3; x = 5
if x > 10:    # (2) x = 5
\end{lstlisting}
These in-lines follow the style proposed by the NExT framework \cite{icml2024next}, which showed that providing such structured and localized execution context can significantly improve the reasoning capabilities of LLMs in program execution understanding tasks.

\begin{table}[t]
\centering
\caption{Prompt templates for {\tool}}
\label{tab:prompt_templates}
\footnotesize
\vspace{-6pt}
\begin{tabular}{p{0.95\linewidth}}
\toprule
\textbf{1. Prompt Template for Test Generation phase} \\
\midrule
Please generate a single pytest test function for the class '\{class\_name\}' based on the provided code under test. Your response should include only one test input. \\
Program under test: \\
---- \\
\{\textcolor{blue}{\texttt{code\_slice}}\} \\
---- \\
The code above does not achieve line: \{\textcolor{orange}{\texttt{target\_line}}\} \\
\textbf{<instructions>} \\
1. Create a new pytest test function\ldots \\
\ldots \\
11. IMPORTANT: Your test method should begin with\ldots \\
Note: If the class under test is an abstract class, do not instantiate it directly in your test. Instead, create a concrete subclass that implements all abstract methods before instantiation, or only test static/class methods without instantiating the abstract class. \\
\textbf{</instructions>} \\
\midrule
\textbf{2. Prompt Template for Test Re-Generation phase} \\
\midrule
Given the following Python program and the function '\{func\_name\}' within the class '\{class\_name\}', your task is to write a test case that executes the specific line of code: \{\textcolor{orange}{\texttt{target\_line}}\}. \\
You are provided with a closely related test case that nearly covers line \{target\_line\}: \\
\{\textcolor{violet}{\texttt{closest\_test}}\} \\
Additionally, here is a cheatsheet containing the gold execution trace for the above example test case. You may use this information to inform your reasoning, but present your analysis as if you derived it independently. \\
\{\textcolor{red}{\texttt{code\_slice\_with\_exec\_inlines}}\} \\
\textbf{<instructions>} \\
Using the provided closest test case, please: \\
1. Analyze why the given test case does not execute the target line. \\
2. Outline a step-by-step strategy to ensure that line is executed. \\
3. Finally, write a new test case that successfully triggers the execution of the target line. \\
Your generated test case should begin with the following template\ldots \\
Please structure your response in two distinct sections: \\
- Enclose step-by-step reasoning within \texttt{`<thinking>`} tags. \\
- Enclose final test case within \texttt{`<answer>`} tags, using backticks for formatting.\\
Important: \\
- If the class under test is abstract, do not instantiate it directly. Instead, create a concrete subclass that implements all abstract methods before instantiation, or only test static/class methods without instantiating the abstract class. \\
\textbf{</instructions>} \\
\bottomrule
\end{tabular}
\end{table}

The structure of the prompts are outlined in Table~\ref{tab:prompt_templates}. By supply\-ing both the logical structure of the code and the actual values~encountered during execution, the LLM is provided with a dual perspective: the static control and data flow from the backward slice, and the dynamic information under a nearly-covering test case. This enriched input guides the LLM to generate a new test case that corrects the previous failure, often by flipping the one condition needed to reach the target line that needs to be covered. This idea is inspired by the Concolic Testing approach~\cite{sen2005cute,godefroid2005dart}, where conditions in a path constraint are flipped to generate new test cases that explore alternative execution paths. However, instead of relying on a SMT solver to find inputs that satisfy the modified constraint, we use the LLM to generate the new test case directly. 

{\color{custom-blue}{After a target statement is covered or skipped, multiple uncovered statements may remain. {\tool} sorts those not-yet-covered statements according to their appearance order in the execution trace and select the first one in that order.
}}






\vspace{-3pt}
\section{Empirical Evaluation}
\label{sec:Experiments}

For evaluation, we seek to answer the following questions:

\vspace{1pt}
\textbf{RQ1. [Effectiveness in Test Generation]} How effective in code coverage is {\tool} in generating test cases in comparison with the baselines \textsc{CoverUp}~\cite{pizzorno2025coverup} and \textsc{CodaMosa}~\cite{lemieux2023codemosa}?

\vspace{1pt}
\textbf{RQ2.} {\bf [Efficiency in Test Generation]} How efficiency is {\tool} in  increasing code coverage during test generation compared to the existing approaches?

\vspace{1pt}
\textbf{RQ3.} [{\bf Ablation Study}] How critical do the components of our method contribute to its performance?

\vspace{1pt}
\textbf{RQ4.} [{\bf Cost Efficiency}] How efficient is {\tool} in coverage-guided test generation in terms of token costs?




We use \texttt{deepseek-v3-0324}~\cite{liu2024deepseek} {\color{custom-blue}{for lower costs}}, which serves as the base LLM for all variants including baselines.

\vspace{1pt}
{\em Datasets.} We conduct our evaluation on the \textsc{CodaMosa} (CM) suite~\cite{lemieux2023codemosa}, a dataset derived from 35 open-source Python projects previously used in the evaluations of BugsInPy and Pynguin. {\color{custom-blue}{It has about 100,000 lines of code across 425 Python modules. We did not use the Pynguin (PY)~\cite{lukas2023py} and MuTAP (MT)~\cite{dakhel2024mutap} datasets because the CM suite is 20$\times$ and 50$\times$ larger than PY (5,000 LOCs, 84 modules) and MT (2,000 LOCs, 163 functions), respectively}}.


\vspace{1pt}
{\em Baselines.} We compare {\tool} with the two baselines: \textsc{coverup}~\cite{pizzorno2025coverup} and \textsc{CodaMosa}~\cite{lemieux2023codemosa}. We reused their core pipelines and replaced GPT-4o services with DeepSeek for cost-efficiency. 


\vspace{1pt}
{\em Procedure.} In the initial test seed phase, we used DeepSeek to generate 10 test cases. In practice, they can be generated by any initial corpus test generation algorithm~\cite{herrera2021seed,pailoor2018moonshine}. During the subsequent test generation phase, which incorporates backward slicing, up to 6 retries are permitted per target line, with each retry representing a single LLM API call aimed at covering that line. In the re-generation phase, any lines that remain uncovered are revisited, allowing up to 5 retries per line. All generated test cases are executed in isolation to prevent state interference and to ensure precise measurement of coverage. For all approaches, we ran for a maximum of 6 hours.


\vspace{1pt}
{\em Metrics.} 
We used three standard metrics for evaluation: {\em line coverage}, {\em branch coverage}, and  combined {\em line+branch} coverage.  line+branch coverage = (\#covered\_lines + \#covered\_branches) / (\#lines + \#branches). Aggregated metrics for each type of coverage are reported as the overall coverage across the entire test suite.

\vspace{-3pt}
\section{Effectiveness of Test Case Generation (RQ1)}
\label{sec:effectiveness}



\begin{figure}[t]
  \centering
  \includegraphics[width=\columnwidth]{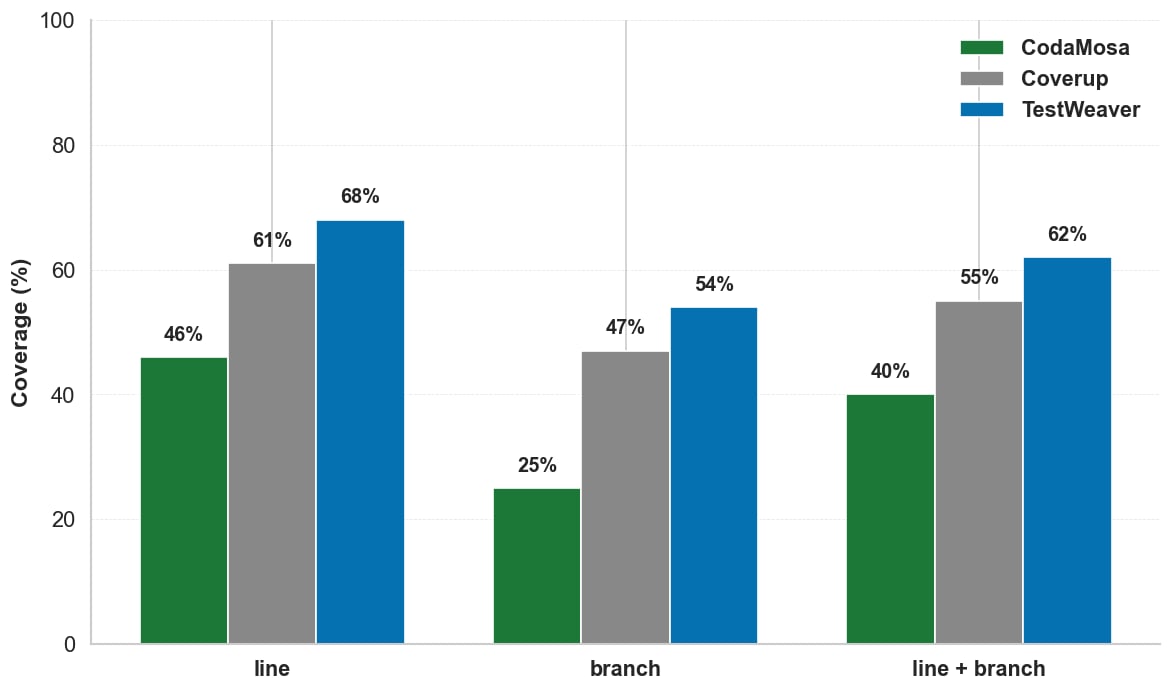}
  \vspace{-21pt}
  \caption{Performance comparison on code coverage (RQ1)} 
  \label{fig:weaver}
\end{figure}

Figure~\ref{fig:weaver} presents the aggregate coverage for the entire benchmark suite, where each value denotes the ratio of the covered lines/branches to the total ones {\em across} the files of 35 projects in our dataset. This applies consistently to line, branch, and line+branch.

Across all settings, as seen in Figure~\ref{fig:weaver}, {\tool} consistently achieves higher coverage than both baseline methods. On the entire dataset, it achieves 68\% line coverage, outperforming \textsc{CoverUp} and \textsc{CodaMosa}, which achieve 61\% and 46\%, respectively. Similarly, for branch coverage, {\tool} reaches 54\%, compared to 47\% and 25\% for the respective baselines. In terms of line+branch coverage, {\tool} achieves 62\%, whereas \textsc{coverUp} and \textsc{CodaMosa} reach 55\% and 40\%, respectively.
In brief, within the same number of hours per repository, {\tool} achieves more line and branch coverages than the two baselines \textsc{CoverUp} and \textsc{CodaMosa}. 


\subsection{Stratification based on Repository Size} 


We aim to study how {\tool} covers the source code in the repositories with different sizes in comparison with the baselines. To this end, we partition the dataset into three groups according to the number of lines of code in each file: Low-\#lines (\(0{-}150\) lines), Mid-\#lines (\(150{-}500\)), and High-\#lines (\(500{-}1,100\)). The total LOCs in each group are non-trivial: Low (18,126), Mid (27,187), and High (10,209).
This allows us to examine how models' performance scales with project size, which is crucial for understanding generalizability.

\begin{table}[t]
\centering
\caption{Stratified coverage result by repository size (RQ1)}
\vspace{-6pt}
\resizebox{0.48\textwidth}{!}{
\begin{tabular}{c l c c c}
\toprule
\textbf{Module} & \textbf{Metric} & \textbf{\textsc{CodaMosa}} & \textbf{\textsc{CoverUp}} & \textbf{\textsc{TestWeaver}} \\
\midrule
\multirow{3}{*}{Low-\#lines} 
& Line            & 50\% & 69\% & \textbf{73\%} \\
& Branch          & 38\% & 66\% & \textbf{69\%} \\
& Line + Branch   & 45\% & 68\% & \textbf{72\%} \\
\midrule
\multirow{3}{*}{Mid-\#lines} 
& Line            & 44\% & 62\% & \textbf{67\%} \\
& Branch          & 27\% & 53\% & \textbf{57\%} \\
& Line + Branch   & 39\% & 59\% & \textbf{64\%} \\
\midrule
\multirow{3}{*}{High-\#lines} 
& Line            & 41\% & 52\% & \textbf{65\%} \\
& Branch          & 23\% & 43\% & \textbf{51\%} \\
& Line + Branch   & 36\% & 47\% & \textbf{57\%} \\
\bottomrule
\end{tabular}
}
\vspace{-3pt}
\label{tab:coverage}
\end{table}

As shown in Table~\ref{tab:coverage}, {\tool} consistently achieves the highest coverage results {\color{custom-blue}{among all the approaches}} across three categories. In low-complexity modules, it outperforms the baselines by margins of 4\%-31\% across all metrics. This performance advantage persists in both medium and high number-of-line groups, where it maintains  a higher coverage from 4\%-13\% over \textsc{CoverUp} and 14\%-30\% over \textsc{CodaMosa}.


All approaches have experienced a moderate decline in code coverage for the entire dataset as the source code has more lines of code, which causes LLMs to navigate through more predictive execution. However, {\tool} remains notably more stable. Its performance degrades only slightly, indicating that the improvements generalize well for large repository sizes.

This robustness arises from our core design choices. First, our backward slicing strategy significantly reduces the size of input to the LLM by retaining only statements relevant to the uncovered targets. To further assess its effectiveness, we collected the {\em lines that \textsc{CoverUp} consistently fail to cover—referred to as difficult lines. {\tool} is able to cover nearly 20\% of these difficult targeted lines}, with the slicing ratios from 13\%-21\%. This demonstrates that backward slicing enhances LLM to pay attention to the most influential statements, mitigating the risk of hallucinations caused by irrelevant or noisy statements and computations—an issue that becomes more severe in large and complex codebases. Second, our constructive feedback mechanism, which selects semantically closest test cases as in-context exemplars, guides the LLM with a more focused context in generating more targeted and logically consistent test cases. By observing concrete parameter values and execution paths, the LLM can better infer which conditions must be altered to reach uncovered lines -- all without actual execution. 

Together, these mechanisms enable {\tool} to scale more effectively with project size, while enhancing precision and reducing erroneous generations—a key advantage over the baselines.


\subsection{Stratification based on Code Complexity}

Sometimes, the number of lines does not reflect well the complexity of the actual execution.
Therefore, we further evaluate TestWeaver’s effectiveness by stratifying all the functions based on their \emph{cyclomatic complexity (CC)}. CC quantifies the structural complexity of a program’s control flow. A CC value above 50 is generally considered \textit{hard and complex}, as it indicates dense branching and deeply nested logic that are challenging for both human reasoning and automated test generation. To assess whether {\tool} remains more effective than the baselines under different complexity levels, we divide the function into four groups with different ranges of CC values: [1--50], [50--100], [100--200], and [200--300]. This stratification allows us to observe how the performance of the approaches scales with increasing code complexity.

\begin{table}[ht]
\centering
\caption{Code Coverage across Functions Stratified by Cyclomatic Complexity (CC) (RQ1).}
\label{tab:cc_stratified_coverage}
\vspace{-6pt}
\resizebox{0.4\textwidth}{!}{
\begin{tabular}{l c c c}
\toprule
\textbf{CC Range} & \textbf{\textsc{CodaMosa}} & \textbf{\textsc{CoverUp}} & \textbf{\tool} \\
\midrule
1--50     & 50\% & 68\% & \textbf{73\%} \\
50--100   & 44\% & 65\% & \textbf{68\%} \\
100--200  & 39\% & 55\% & \textbf{63\%} \\
200--300  & 45\% & 57\% & \textbf{66\%} \\
\bottomrule
\end{tabular}
}
\end{table}

As seen in Table~\ref{tab:cc_stratified_coverage}, \emph{\tool consistently outperforms both \textsc{CoverUp} and \textsc{CodaMosa} across all complexity groups}. The performance gap becomes more pronounced as complexity increases, highlighting \tool’s robustness in handling intricate control flows. This improvement stems from its program slicing and execution-inlining strategies, which help the LLM focus on relevant control and data dependencies—reducing distractions and hallucinations even in complex scenarios. In our ablation study (Section~\ref{sec:ablation}), we will show that with our execution-inlining and closest test retrieval, {\tool} handles better with intricate code.

\section{Efficiency of Test Case Generation (RQ2)}
\label{sec:efficiency}

\begin{figure}[t]
  \centering
  \includegraphics[width=\columnwidth]{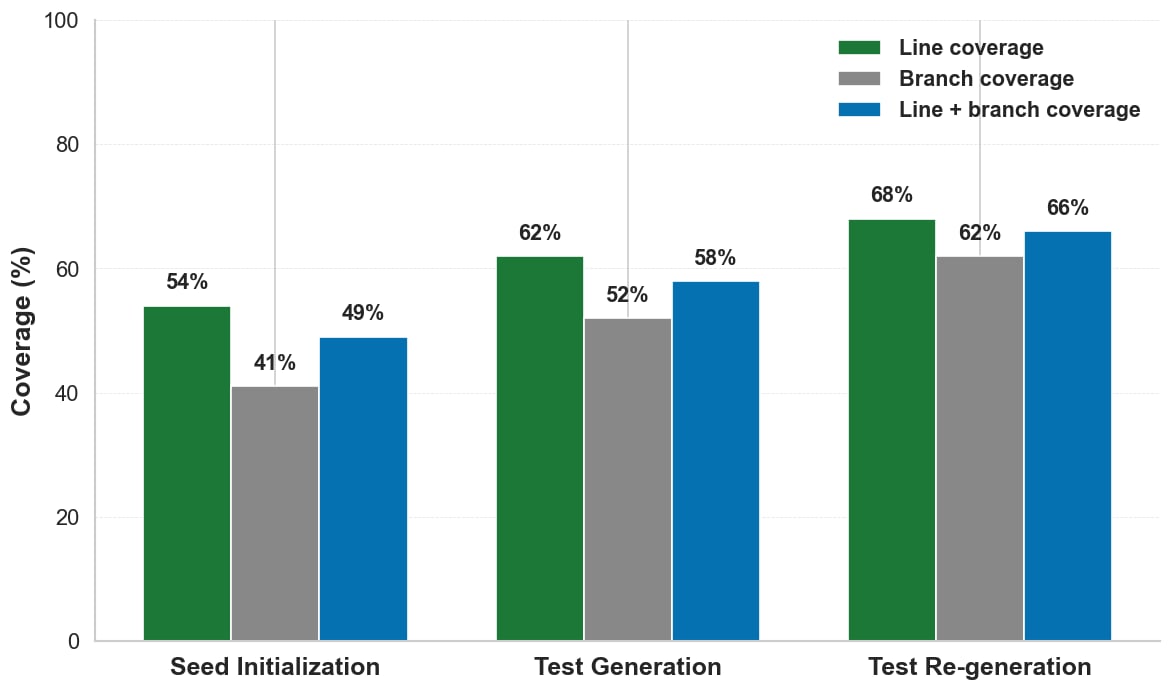}
  \vspace{-18pt}
  \caption{ Coverage improvements across three phases of {\tool} on the CM suite (RQ2).
  }
  \label{fig:coverage-increase}
\end{figure}


In this experiment, we aim to investigate how code coverage evolves throughout three phases of test generation in {\tool}: (1) \textit{Seed Initialization}, (2) \textit{Test Generation}, and (3) \textit{Test Re-Generation}.

As seen in Figure~\ref{fig:coverage-increase}, the coverage starts at 54\% after the initial phase and increases to 62\% during the second phase, highlighting the effectiveness of our backward slicing strategy, trimming down the irrelevant statements to help the LLMs focus on the important ones. By limiting the LLM’s attention to only the code statements relevant to the uncovered targets, slicing reduces the input size to just 69–78\% of the original code, resulting in more focused and effective prompt construction. 
This result confirms our Key Idea~1 in our design listed in Section~\ref{sec:key-idea-1}.
In the third phase, coverage rises further to 68\%. While slicing narrows the context, LLMs may still struggle with complex control flows or variable interactions. To address this, {\tool} incorporates our feedback mechanism, specifically provides additional guidance including the ``closest'' test cases and inlined execution information. These two feedbacks help the LLM better infer execution semantics and reduce hallucination.

\begin{figure}[t]
  \centering
  \includegraphics[width=\columnwidth]{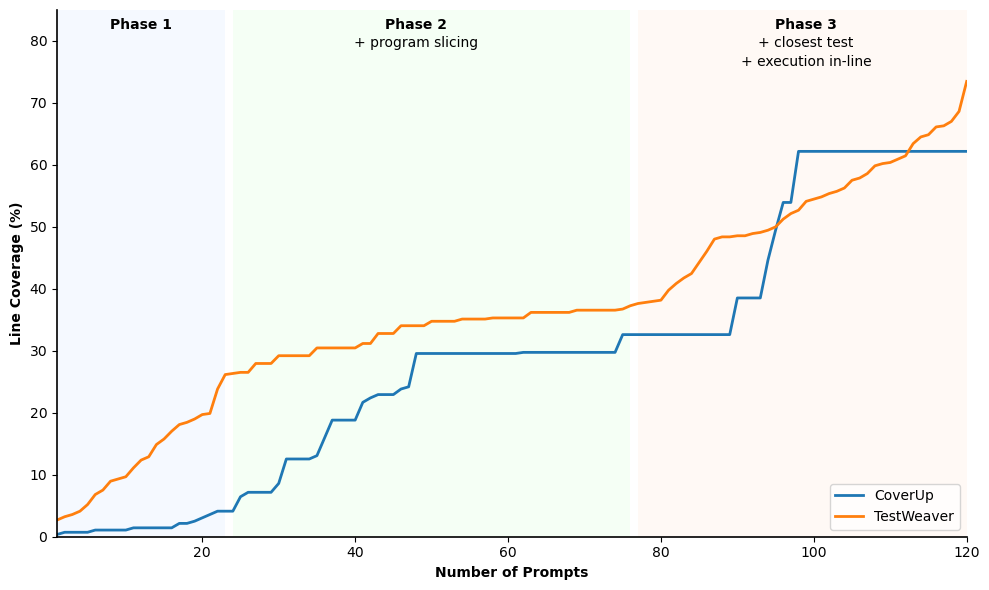}
  \vspace{-18pt}
  \caption{Line coverage progression across 120 prompts for \texttt{typesystem.fields} module. Colored areas show phases (RQ2)}
  \label{fig:prompt_progression}
\end{figure}

\paragraph{Comparative Analysis}
Notably, our tool remains more effective even on highly complex modules in the CM suite. Let us use a case study on the module \code{typesystem.fields}, exceeding 550 lines with a cyclomatic complexity over 350. We further illustrate this improvement by comparing {\em per-prompt coverage progression} between {\tool} and the best baseline, \textsc{CoverUp} (Figure~\ref{fig:prompt_progression}). The X-axis shows the number of prompts, while the Y-axis shows the line coverage in percentage.

To compare with the best baseline \textsc{CoverUp} (Figure~\ref{fig:prompt_progression}), {\tool} consistently achieves higher line coverage with each prompt, while \textsc{CoverUp} frequently encounters coverage plateaus. Notably, its longest plateau persists for about 30 consecutive LLM retries. After reaching its maximum coverage at the 98th prompt, \textsc{CoverUp} stalls at 63\% coverage, even after more than 20 additional retries. In contrast, {\tool} continues to make progress: it surpasses 63\% at the 112th prompt and reaches 75\% by the 120th prompt.

A closer look at the results shows that this improvement comes from (1) the reduced number of statements the LLM must handle thanks to backward slicing in both the initial and test generation phases, and (2) the {\em constructive feedback strategies} applied during retries in the final test re-generation phase. 

\textsc{CoverUp} uses a {\em more passive retry strategy, simply listing all uncovered lines in a prompt to the LLM for another attempt}. This provides no guidance on how to cover those lines, which may even require conflicting execution paths within a single test case. As a result, the LLM lacks critical execution context and is forced to guess, often repeating the same mistakes. In Figure~\ref{fig:prompt_progression}, {\em in the total of 120 prompts to get to 63\% of line coverage, 69 re-prompts to the LLM from \textsc{CoverUp} did not result any new line covered. The top-3 coverage plateaus in that example lasted 14, 20, and 30 prompts}.

\definecolor{darkgreen}{rgb}{0.0, 0.5, 0.0}
\begin{table}[t]
  \centering
  \caption{Comparison of retry strategies in \textsc{CoverUp} and constructive feedback mechanisms in {\tool} (RQ2).}
  \label{tab:coverage-strategy}
  \vspace{-6pt}
  \resizebox{\columnwidth}{!}{
  \begin{tabular}{p{3.2cm} p{3.3cm} p{3.5cm}}
    \toprule
    \textbf{Aspect} & \textbf{\textsc{CoverUp}} & \textbf{\tool} \\
    \midrule
    Constructive feedback & \textcolor{red}{Listing all non-covered lines in prompt} & \textcolor{darkgreen}{Providing "closest" test case and inline explanation} \\
    Execution awareness & \textcolor{red}{Ignores runtime execution data} & \textcolor{darkgreen}{Grounded in actual execution via inline traces} \\
    Coverage progression & \textcolor{red}{More coverage plateaus due to weak feedback} & \textcolor{darkgreen}{Consistently improves with targeted guidance} \\
    Code reduction strategy & \textcolor{red}{Function-level segmentation only} & \textcolor{darkgreen}{Line-level backward slicing to exclude irrelevant code} \\
    \bottomrule
  \end{tabular}
  }
\end{table}

In contrast, {\tool} offers more constructive feedback by retrieving the "closest" test case and embedding execution traces directly into the prompt.
This richer, execution-aware feedback helps {\tool} recover from failures and make continuous progress. Additionally, unlike \textsc{CoverUp}, which segments code only at the function level, {\tool} performs fine-grained program slicing to exclude lines that do not affect the target line’s execution. This reduces unnecessary context and minimizes hallucination.
Consequently, while \textsc{CoverUp} often hits an early coverage ceiling due to its shallow feedback loop and coarse context selection, {\tool} reduces this plateau by providing precisely the information needed to resolve each uncovered line. In Figure~\ref{fig:prompt_progression}, {\tool} has only 15 times that retries did not result in any increasing coverage. Table~\ref{tab:coverage-strategy} summarizes the key improvements of {\tool} over the baseline, with the maximum coverage plateaus of 5 prompts.





\section{Ablation Study (RQ3)}
\label{sec:ablation}

{\tool} has three key components: (1) backward slicing in the test generation phase, (2) ``closest'' test case selection, and (3) execution inlining in the test re-generation phase. In this experiment, we study their contributions by disabling each of them.


\vspace{1pt}
{\bf \em Backward Slicing}. Table~\ref{tab:phase2-testgen_ablation} presents the effect of disabling backward slicing. While program slicing improves coverage by 4\%--7\% overall, this gain reflects its ability to assist the LLM in reducing hallucination. We further analyze {\em the cases where test generation fails to cover the target line without slicing but succeeds with slicing; in these, slicing eliminates 16\%–25\% of the original code per file, allowing the LLM to focus on fewer, more targeted paths}. {\color{custom-blue}{This result corroborates our Key Idea 1 on using program slicing.}} 


\begin{table}[t]
\centering
\small
\caption{Ablation on test generation: overall coverage obtained on the CM Suite (more is better) (RQ3)}
\label{tab:phase2-testgen_ablation}
\vspace{-6pt}
\begin{tabular}{l@{\hskip 3pt}l@{\hskip 3pt}l@{\hskip 3pt}c}
\toprule
\textbf{\% Coverage} & \textbf{Line} & \textbf{Branch} & \textbf{Line + Branch} \\
\midrule
\tool & 
\multicolumn{1}{l}{\textbf{62\%}} & 
\multicolumn{1}{l}{\textbf{52\%}} & 
\multicolumn{1}{l}{\textbf{58\%}} \\
{\small -- w/o Slicing} & 
\multicolumn{1}{l}{58\% \textcolor{red}{{(-4\%)}}} & 
\multicolumn{1}{l}{45\% \textcolor{red}{{(-7\%)}}} & 
\multicolumn{1}{l}{52\% \textcolor{red}{{(-6\%)}}} \\
\bottomrule
\end{tabular}
\end{table}

\begin{table}[t]
\centering
\small
\setlength{\tabcolsep}{1pt} 
\caption{Ablation for test re-generation: overall coverage obtained on the CM Suite (more is better) (RQ3)}
\label{tab:phase3-test_regen_ablation-1}
\vspace{-6pt}
\renewcommand{\arraystretch}{1.0} 
\begin{tabular}{l@{\hskip 3pt}l@{\hskip 3pt}l@{\hskip 3pt}c}
\toprule
\textbf{\% Coverage} & \textbf{Line} & \textbf{Branch} & \textbf{Line + Branch} \\
\midrule
\tool & 
\textbf{68\%} & 
\textbf{62\%} & 
\textbf{66\%} \\
{\small -- {\color{custom-blue}{w/o Execution Inlining}}} & 
63\% \textcolor{red}{{(-5\%)}} & 
56\% \textcolor{red}{{(-6\%)}} & 
61\% \textcolor{red}{{(-5\%)}} \\
{\small -- w/o Closest Test Sel.} & 
62\% \textcolor{red}{{(-6\%)}} & 
52\% \textcolor{red}{{(-10\%)}} & 
58\% \textcolor{red}{{(-8\%)}} \\
\bottomrule
\end{tabular}
\end{table}



\vspace{2pt}
{\bf \em Closest Test Selection and Execution Inlining.} Table~\ref{tab:phase3-test_regen_ablation-1} illustrates the impact of the closest test selection and in-line execution components. The results show that removing either component consistently degrades coverage. Among them, removing the closest test mechanism leads to the most noticeable drop—between 6\%–-10\% across coverage metrics—while excluding in-line execution results in a 5\%--6\% decrease. Notably, the two factors are complementary: leveraging a closest test case and adding execution information provide the LLM with dynamic signals, leading to {\tool}'s most effectiveness in generating test cases to cover target lines. {\color{custom-blue}{This result corroborates our Key Ideas 2--3 in leveraging the closest test case and execution data to generate better test cases.}}



\vspace{2pt}
{\bf \em Test Selection Strategies.}
{\color{custom-blue}{
Table~\ref{tab:phase3-test_regen_ablation-2} presents an ablation study evaluating the importance of the closest test case selection strategy in the test re-generation phase.
In the standard setting, {\tool} identifies the ``closest'' test case to the target line (yet not covering it), and uses its execution trace to annotate the program slice with execution in-lines. This configuration achieved 68\% line coverage for entire dataset. In contrast, replacing the closest test case with a {\em randomly selected test case} from the current test suite reduced coverage to 65\%, indicating that {\em an arbitrary execution context provides weaker guidance to the LLM, leading to lower code~coverage}. 

A second variant incorporated execution traces from $k$ random test cases ($k$=5) into the prompt, without prioritizing proximity to the target line. This multi-trace setup achieved 64\% coverage—lower than both the random baseline and the closest-test case strategy. This suggests that execution in-lines derived from the ``closest'' test case play a critical role in guiding the LLM in effective test generation to cover the target line. This also confirms the effectiveness of our feedback with test cases and ``closest'' test selection algorithm.
}}


\begin{table}[t]
\centering
\small
\caption{{\color{custom-blue}{Ablation for ``closest'' test selection strategies (RQ3).}}}
\label{tab:phase3-test_regen_ablation-2}
\vspace{-9pt}
\renewcommand{\arraystretch}{1} 

\begin{tabular*}{0.6\linewidth}{@{\extracolsep{\fill}} l r}
\toprule
\textbf{\% Coverage} & \textbf{Line} \\
\midrule
Closest Test Case    & \textbf{68\%} \\
Random Test Case      & {\color{red}(-3\%)} 65\% \\
$k$ Random Test Cases    & {\color{red}(-4\%)} 64\% \\
\bottomrule
\end{tabular*}
\end{table}

\section{Efficiency on Token Costs (RQ4)}

\begin{figure}[t]
    \centering
    \includegraphics[width=\columnwidth]{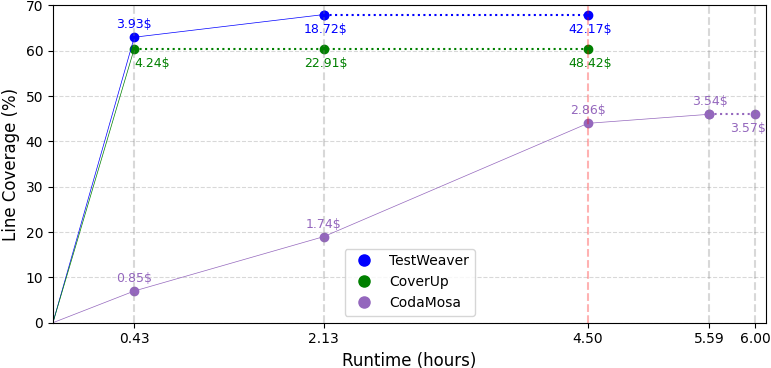}
    \vspace{-21pt}
    \caption{Coverage progression over running time with token cost in USD. 
    A dotted line indicates coverage plateau~after last saturation, where no further improvement occurs (RQ4).} 
    \label{fig:runtime_cost}
\end{figure}




{\color{custom-blue}{
In this experiment, we aim to assess {\tool}'s time and cost efficiency in comparison with those of the two baselines. Since \textsc{CoverUp}'s authors set the running time of 4 hours in their experiments, we executed {\tool} and \textsc{CoverUp} for 4.5 hours for a fair comparison. Since \textsc{CodaMosa}'s coverage increases more slowly, we let it run up to 6 hours to observe coverage changes.

To assess time and cost efficiency with regard to coverage, along the execution of a tool, we tracked {\em the \underline{last} saturation point}--{\em defined as the time where the further (remaining) retries yield no additional line coverage}. The last saturation point corresponds to the last coverage plateau for a tool's run, allowing us to assess {\em a tool’s peak coverage and the  earliest time} at which it is reached.
We recorded the following metrics for comparison: the current running time, the (token) cost in USD for LLM usages, and the achieved coverage. We recorded those metrics at the last saturation point and at the run end for a tool, and the respective values of the other tools.

\vspace{2pt}
\noindent {\bf \em Last Saturation Point and Coverage.} As seen in Figure~\ref{fig:runtime_cost}, at~the~ending time, {\tool} achieves {\em the highest coverage for the~entire dataset} (68\%) (compared to \textsc{CoverUp} (61\%) and  \textsc{CodaMosa} (46\%)). 

Compared to \textsc{CoverUp}, which hits its last saturation early at 0.43 hour, {\tool} has reached 62\% coverage by that time—at lower cost. At its last saturation (2.13), {\tool} reaches higher coverage (68\%), and has a shorter last plateau (i.e., no further gains, blue vs. green dotted lines). In contrast, \textsc{CodaMosa} has a later last saturation point (5.59 hours) than {\tool}, yet attains only 46\% coverage, even with more time. {\em This shows {\tool}'s better coverage performance at all checkpoints (last saturation and ending).}




This plot {\em does not have the coverage plateaus from a tool before the last saturation point} since it aims to show only  efficiency with regard to the {\bf \em total coverage} of all  repositories in our entire dataset. In our experiment, at any time before the last saturation point, there exists {\em one or more repositories in the dataset with increasing coverages}, leading to {\em increasing total coverage}. Thus, {\em this timeseries plot for the total coverage on the entire dataset does not contain~the coverage plateaus before the last saturation}. Moreover, large repositories need more running time. 
Thus, for coverage plateaus, in Section~\ref{sec:efficiency}, we showed them for a complex repository on the prompt basis. The coverage plateaus for other repositories~\cite{testweaver-website} followed similar trends as in Figure~\ref{fig:prompt_progression}. In brief, this plot with last saturations shows the time to the peak coverage at the last plateau, rather than all plateaus.


\vspace{2pt}
\noindent {\bf \em Cost and Coverage}. Despite having lower token costs due to~no constructive feedback as in {\tool}, \textsc{CodaMosa}'s coverage progresses more slowly, requiring the entire 6 hours to reach its final coverage of only 46\%, compared to {\tool}'s 2.13 hours with 68\%. At any checkpoint, \textsc{CodaMosa}'s coverage is {\em much lower~than} {\tool}'s. Thus, {\em the tradeoff—higher token cost than \textsc{CodaMosa} in exchange for superior coverage— is justified for {\tool}}.




At 0.43 hour, {\tool} already attains 62\% coverage with \$3.93 compared to \textsc{CoverUp}'s only 61\% with \$4.24. By 2.13~hours, it reaches 68\% coverage at a cost of \$18.72, while \textsc{CoverUp} remains stuck at 61\% even after running for the same duration at a higher cost (\$22.91). Extending \textsc{CoverUp}'s running to 4.5 hours results in no further gain, as coverage remains at 61\% with a cost of \$48.42. {\tool}'s lower cost compared to that of \textsc{CoverUp} stems from its shorter, sliced prompts, which reduce token usage while keeping the LLM focused on relevant execution paths. In brief, {\tool} advances \textsc{CodaMosa} and \textsc{CoverUp} in terms of coverage and cost.

}}

\vspace{-3pt}
\section{\bf Limitations and Threats to Validity}


\noindent {\em \underline{Limitations}}: First, our re-implementation of backward slicing may omit essential contexts from dependent classes, potentially leading to incorrect type inference for method parameters. Second, similar to \textsc{CoverUp}, LLMs sometimes generate un-executable test code. We use the same program analysis module in \textsc{CoverUp} to resolve the compilation errors. Third, in cases where the target line is hard to reach, no closest test case may be found—leaving the LLM without a strong exemplar to guide generation, hindering effectiveness. Finally, after a fixed number of attempts with "closest" test case and execution inlines to cover a line, we move on to the next line.

\noindent {\em \underline{External Validity}}: Our chosen benchmarks might not be representative, but they have been used in the existing work, enabling a fair comparison. We evaluated {\tool} with \code{deepseek-v3-0324}. The results may vary for other LLMs.  We evaluated {\tool} only on Python. More experiments are needed for other languages. 

\noindent {\em \underline{Internal Validity}}: The third-party tools for dependence analysis or the compiler, may introduce errors. While the code in our dataset may be publicly available, very few associated test cases are accessible. Furthermore, the available inputs are limited in scope 
and domains. Due to the lack of comprehensive source code coverage and the infeasibility of exhaustively executing all possible inputs, {\em the risk of data leakage is minimal}. As such, the soundness of our evaluation remains unaffected. The network traffics, latency, server loads when interacting with LLMs affected the number of API calls to the LLMs. Finally, potential limitations in our re-implementation of a dynamic backward slicing algorithm can introduce bugs.

\noindent {\em \underline{Construct Validity}}: Varied prompts may lead to varied input distributions, potentially affecting results. To address this, we define the prompts with well-specified~structure as shown in Table~\ref{tab:prompt_templates}.

\vspace{-6pt}
\section{Related Work}
\label{sec:related}

{\tool} is closely related to \textsc{CoverUp}~\cite{pizzorno2025coverup}, which combines coverage analysis, code context, and iterative feedback in prompting LLMs 
to improve code coverage. However, {\tool} introduces key advancements. First, {\tool} provides more focused context to the LLM by leveraging backward slicing from the target line, whereas CoverUp supplies function-level code segments.
Second, CoverUp only considers passive feedbacks with all the uncovered lines and retries the prompting without any helpful guidance. 
In contrast, {\tool} augments the prompt with execution in-lines that expose relevant variables and program state, thereby helping LLMs 
reach the target line. Finally, CoverUp uses PA to fix the syntactically incorrect generated test code. We leverage PA to improve both incorrectly generated tests and LLMs' test generation.


A number of LLM-based test generation techniques have been proposed~\cite{10440574}. \textsc{CodaMosa}~\cite{lemieux2023codemosa} addresses the coverage plateau by prompting an LLM for a new test case when the search-based test generation process stalls. 
Bareiß {\em et al.}~\cite{bareiß2022codegenerationtoolsalmost} evaluate the performance of Codex on Java unit test generation. Their prompts include the method signature, an example test case, and the method body, while discarding any non-compiling outputs. Vikram {\em et al.}~\cite{vikram2024largelanguagemodelswrite} investigate the use of modern LLMs to automatically synthesize property-based tests using API documentation. TiCoder~\cite{10606356,lahiri2023interactivecodegenerationtestdriven} prompts LLMs to generate tests from textual descriptions of intended code functionality, enabling test-driven development. TestPilot~\cite{10329992} generates JavaScript unit tests by prompting an LLM with the function under test, its documentation, and related usage examples.  ChatUniTest~\cite{chen2024chatunitest} focuses on Java unit test generation by prompting LLMs with source code; it also includes mechanisms to repair non-compiling tests. TestEval~\cite{wang-etal-2025-testeval} benchmarks LLMs on LeetCode problems, while TestGenEval~\cite{jain2025testgeneval} is a project-level suite focused on testing goals like exception handling and boundary~coverage.

Fuzz4All~\cite{fuzz4all} uses two LLMs in tandem to perform fuzz testing across programming languages. SymPrompt~\cite{ryan2024codeaware} aims to generate tests that reach hard-to-cover code regions via prompting LLMs to focus on their path constraints.
TestGen-LLM~\cite{testgen-llm}
prompts the LLM with the class under test—and optionally the existing test class—to produce additional coverage-enhancing test cases. MuTAP~\cite{DAKHEL2024107468} focuses on mutation testing in Python. It prompts the LLM to generate an initial test, performs mutation testing, and prompts again to generate new~assertions for surviving mutants, which are added to the test suite.
ExLong~\cite{zhang2025exlong} is an instruction fine-tuned LLM built on CodeLlama to reason on traces to methods containing throw statements or guard
expressions for exceptional behavior tests.



{\color{custom-blue}{
Several approaches assist the LLMs in reasoning about the dynamic program behaviors~\cite{yi2025blended,bieber2019learning,liu2023code,la2024code,lyu2024large,tufano2023predicting,dhulipala2024planning,icse25-orca,liu2024codemind, chi-etal-2025-visualcoder,ding2024traced,icml2024next}. 
IPA-GNN~\cite{bieber2019learning} is a GNN variant tailored toward leverage CFGs to predict program execution. 
VisualCoder~\cite{chi-etal-2025-visualcoder} leverages multi-modal Chain-of-Thought (CoT) reasoning to improve the execution prediction using CFG expressed in image and text representation.
\textsc{PredEx}~\cite{yi2025blended} combines LLMs with predictive backward slicing to predict the next executed statement.
However, IPA-GNN, VisualCoder, and PredEx predict program execution traces and variable values from a given input without actual execution. In contrast, {\tool} generates an input test case to cover a target statement. 

La Malfa {\em et al.}~\cite{la2024code} propose a benchmark to assess LLMs in simulating complex code. Their Chain of Simulation instructs LLMs to simulate program execution following the computation pattern of compilers. Similarly, Orca~\cite{icse25-orca} leverages ReAct planning~\cite{yao2023react}
to instruct LLMs to predict program execution to detect runtime errors. CodePilot~\cite{dhulipala2024planning} utilizes CoT planning to predict code coverage for a given input, improving over simple prompting in Tufano {\em et al.}~\cite{tufano2023predicting}. Liu {\em et al.}~\cite{liu2023code} propose to pre-train LLMs with execution data. Lyu {\em et al.}~\cite{lyu2024large} introduce Iterative Instruction Prompting (IIP) to improve over CoT in code execution prediction. TRACED~\cite{ding2024traced} is an execution-aware pre-training strategy for source code with a combination of source code, executable inputs, and execution traces. These planning/prompting frameworks and execution-aware pre-train strategies can be used to enhance our prompting in the test case re-generation phase because they can help the LLM learn the execution nuances.
Inspired by NeXT~\cite{icml2024next}, we used execution inlining to provide focused context for execution reasoning.
CodeMind~\cite{liu2024codemind} aims to gauge LLMs' reasoning via inductive reasoning. 
Deep learning has been used to infer inputs from desired outputs via program synthesis, input-output reasoning~\cite{zaremba2014learning,mammadov2024learning, chen2021latent, parisotto2016neuro, 10.5555/3305381.3305484}.

}}

\section{Conclusion and Implications}

\paragraph {\bf \em Novelty} This paper introduced {\tool}, a novel LLM-based framework that overcomes the limitations of existing test generation approaches by integrating execution-aware constructive feedback  into the test generation loop. Unlike prior work that relies solely on passive retrying in prompting, {\tool} brings three key innovations: backward slicing to focus the model’s attention on execution-relevant code, retrieval of control-flow-"closest" test cases to provide context-efficient guidance, and execution in-line annotations to help LLMs reason more effectively about program behavior. These techniques jointly address the long-standing issue of coverage plateau by equipping LLMs with targeted, execution-aware inputs that reduce redundancy and improve path exploration. 

 
\vspace{2pt}
{\bf \em Implications.} {\tool} opens new avenues for future research at the intersection of program analysis and large language models in software testing. By demonstrating that lightweight static and dynamic analyses can meaningfully augment LLM reasoning, {\tool} encourages a shift from purely prompt-based techniques to hybrid methods that integrate symbolic information and execution context. This approach could inspire the development of smarter, context-aware LLM agents not only for regression testing but also for broader tasks such as debugging, program repair, and test oracle inference. Furthermore, {\tool}’s notion of "closeness" between execution paths may serve as a foundation for designing novel guidance strategies in other test generation paradigms, such as symbolic or concolic testing. In practical settings, {\tool}'s ability to generate high-coverage test cases could significantly reduce bugs and accelerate continuous integration.

\vspace{2pt}
\noindent {\bf \em Data Availability}. https://github.com/FSoft-AI4Code/TestWeaver



\section*{Acknowledgments}
This work was supported in part by the US National Science Foundation (NSF) grant CNS-2120386.



\newpage


\bibliographystyle{ACM-Reference-Format}
\bibliography{references,references-predex,references-predex1}



\end{document}